\documentclass[12pt]{article}
\makeatletter
\@addtoreset{equation}{section}
\makeatother

\begin{document}
\vspace*{0.5cm}

\begin{center}
{\bf THE HIGGS SECTOR IN THE MINIMAL 3-3-1 MODEL WITH THE MOST
GENERAL LEPTON-NUMBER CONSERVING POTENTIAL} \vspace*{1cm}

\bigskip

{\bf Nguyen Tuan Anh $^a$}, ~ {\bf Nguyen Anh Ky $^{a,b}$} and {\bf
Hoang Ngoc Long $^{a, c}$} \vspace*{0.5cm}

\medskip

$^a$ {\it Institute of Physics, NCNST, P. O. Box 429, Bo Ho, Hanoi
10000,
Vietnam}\\
$^b$ {\it Department of Physics, Chuo University, Bunkyo-ku,
Tokyo 112-8551, Japan}\\

$^c$ {\it LAPTH, B.P. 110, F--74941,
Annecy--le--Vieux Cedex,
France}\\
\vspace*{5mm}

\bigskip \bigskip \bigskip \bigskip

\end{center}

\begin{abstract}
  The Higgs sector of the minimal 3 - 3 - 1 model with three triplets
and one sextet is investigated in detail under the most general
lepton--number conserving potential. The mass spectra and
multiplet decompostion structure are explicitly given in
a systematic order and a transparent way allowing they to
be easily checked and used in further investigations. A
previously arising problem of inconsistent signs of $f_{2}$
is also automatically solved.
\end{abstract}
\vskip .7 true cm

PACS numbers: 11.15.Ex, 12.60.Fr, 14.80.Cp.\\


\section{Introduction}

  The standard model (SM) combining the
Glashow--Weinberg--Salam (GWS) model with
the QCD under the gauge group $SU_{C}(3)\otimes
SU_{L}(2)\otimes U_{N}(1)$ is one
of the greatest achievements of physics in the 20th
century. Many predictions of
the SM have been confirmed by various experiments.
However,
this model works well
only at the energy range below 200 GeV and gradually
losses its prediction power
at higher energies. Therefore, any extension of the SM
to fit the theory with the
higher energy phenomenology is needed. In addition,
the observation of the Higgs
bosons which play a central role in symmetry breaking is
still an open problem. Hence the mechanism for electroweak
symmetry breaking is,
in some way, still a mystery. The scalar sector has been
throughly studied not
only in the framework of the SM but also in its various
extensions including the
so--called 3 - 3 - 1 models based on the
$SU_{C}(3)\otimes SU_{L}(3) \otimes U_{N}(1)$
 gauge group [1--6]. The later models have the following
intriguing features:
firstly, the models are anomaly free only if the number of
families N is a
multiple of three. Further, the condition of QCD asymptotic
freedom valid only
for the number of quark families less than five, leads to
 N equal to 3. Secondly,
 the Peccei--Quinn (PQ) ~\cite{pq} symmetry -- a solution
 of the strong CP problem
 naturally occurs in the 3 - 3 - 1 models~\cite{pal}.
 It is worth mentioning that the
 implementation of the PQ symmetry is usually possible only
 at a classical level
 (broken by quantum corrections through instanton effects)
 and there has been a
 number of attempts to find models solving the strong CP question.
 In these 3 - 3 - 1
 models the PQ symmetry following from the gauge invariant
 Lagrangian does not
 have to be imposed. The third interesting feature is that
 one of the quark
 families is treated differently from the other two.
 This could lead to a natural
 explanation for the unbalancing heavy top quarks, deviations
 of $A_b$ from the SM
 prediction, etc. Additionally, the models predict not very
 high new mass scales,
 at the order of a few TeV only \cite{dng}.\par
\bigskip

  Recently, the scalar sector of the minimal 3 - 3 - 1 model was in
 detail studied in ~\cite{ton} and \cite{tkl}. There, three Higgs
 triplets were  firstly  analysed and then the sextet was added in a
 further consideration.
It was also shown in \cite{tkl} that the potential used
in \cite{ton} leaded to
inconsistent results and therefore it should be further
modified or replaced by some more relevant potential.
Another precise
investigation on the model with a new potential is,
in our opinion,
really interesting and necessary.
Following the previous paper \cite{tkl} the present paper
is devoted
to such an investigation. Here, instead of the potential
in \cite{ton,tkl} the
most general gauge--invariant postential conserving lepton
numbers \cite{tj} is used. With the latter potential
the scalar sector of the 3 - 3 - 1
model is investigated again at tree--level.
The multiplet decomposition structure
remains the same as in \cite{tkl} but the masses
of most of the
scalars get corrections and the problem with inconsistent
signs of $f_2$ is automatically solved. We emphasize that
the above mentioned potential was also considered in
\cite{tj,plei}, but only mass matrices~\cite{tj} and
some their
eigenvalues~\cite{plei} were presented.\bigskip

   The paper is organized as follows. The Higgs potentials,
constraint equations
and main notations are presented in section {\bf 2}, while
the characteristic
equations are solved in section {\bf 3} where the multiplet
decompositions and
mass spectra of the Higgs sector are given.
Our results are summarized in the
last section, section {\bf 4}.

\section{The Higgs contents and scalar potentials}

 Presently, the minimal 3 - 3 - 1 models are considered
with three Higgs triplets ~\cite{ppf}
\begin{equation}
\eta =\left(
\begin{array}{l}
\eta ^{o} \\
\eta _{_{1}}^{-} \\
\eta _{_{2}}^{+}
\end{array}
\right) \sim (1,3,0),\ \rho =\left(
\begin{array}{l}
\rho ^{+} \\
\rho ^{o} \\
\rho ^{++}
\end{array}
\right) \sim (1,3,1),\ \chi =\left(
\begin{array}{l}
\chi ^{-} \\
\chi ^{--} \\
\chi ^{o}
\end{array}
\right) \sim (1,3,-1),
\label{triplets}
\end{equation}
and one Higgs sextet \cite{rf}
\begin{equation}
S=\left(
\begin{array}{ccc}
\sigma _{_{1}}^{o} & s_{_{2}}^{+}/\sqrt{2} & s_{_{1}}^{-}/\sqrt{2} \\
s_{_{2}}^{+}/\sqrt{2} & s_{_{1}}^{++} & \sigma _{_{2}}^{o}/\sqrt{2} \\
s_{_{1}}^{-}/\sqrt{2} & \sigma _{_{2}}^{o}/\sqrt{2} & s_{_{2}}^{--}
\end{array}
\right) \sim (1,6^{\ast },0).
\label{sextet}
\end{equation}
The latter is needed in order to give masses to all leptons.
In \cite{ton, tkl}, the scalar sector of the minimal
3 - 3 - 1 models is investigated by using the potential
\begin{eqnarray}
V_{_S}(\eta ,\rho ,\chi ,S) &=&V_{_T}(\eta ,\rho ,\chi )+\mu
_{_4}^2{Tr}(S^{\dagger }S)+\lambda _{_{10}}{Tr}^2
(S^{\dagger }S)+
\lambda
_{_{11}}{Tr}[(S^{\dagger }S)^2]  \nonumber \\
&&+[\lambda _{_{12}}\eta ^{\dagger }\eta +\lambda _{_{13}}
\rho ^{\dagger
}\rho +\lambda _{_{14}}\chi ^{\dagger }\chi ]
{Tr}(S^{\dagger }S)
\nonumber
+2f_{_2}\left( \rho ^TS\chi +h.c.\right),
\label{sympoten}
\end{eqnarray}
with
\begin{eqnarray}
V_{_T}(\eta ,\rho ,\chi ) &=&\mu _{_1}^2\eta ^{\dagger }\eta +\mu
_{_2}^2\rho ^{\dagger }\rho +\mu _{_3}^2\chi ^{\dagger }\chi
+\lambda _{_1}(\eta ^{\dagger }\eta )^2+\lambda _{_2}(\rho
^{\dagger }\rho )^2+\lambda _{_3}(\chi ^{\dagger }\chi )^2
\nonumber \\ &&+\lambda _{_4}(\eta ^{\dagger }\eta )(\rho
^{\dagger } \rho )+\lambda _{_5}(\chi ^{\dagger }\chi )(\eta
^{\dagger }\eta )+ \lambda _{_6}(\rho ^{\dagger }\rho )(\chi
^{\dagger }\chi )  \nonumber
+\lambda _{_7}(\rho ^{\dagger }\eta )(\eta ^{\dagger }
\rho )\\
&&+\lambda
_{_8}(\chi ^{\dagger }\eta )(\eta ^{\dagger }\chi )+
\lambda _{_9}(\rho
^{\dagger }\chi )(\chi ^{\dagger }\rho )
+\sqrt{2}f_{_1}\left( \varepsilon ^{ijk}\eta _{_i}
\rho _{_j}\chi_{_k}+h.c.\right),
\label{tpoten}
\end{eqnarray}
where $\mu $'s are mass parameters and the
coefficients $f_{_1}$
and $f_{_2}$ have dimensions of mass, while
$\lambda $'s are
dimensionless. Unfortunately, as shown in \cite{tkl},
this potential
(which is not most general) leads to nonlogical
results as $f_2$
cannot take a consistent sign. Analysing the
results obtained, we
conclude that we need a wider potential in order to
give necessary
corrections to those masses showing contradict signs
of $f_2$. On the other hand,
we suggest that the potential needed should guarantee
the conservation of the
lepton numbers and the continuous symmetry not
higher than $SU(3)\times U(1)$.
Fortunately, there exist such potentials.
The most general potential satisfying
our requirements is \cite{tj}
\begin{eqnarray}
V_{_E}(\eta ,\rho ,\chi ,S) &=&V_{_S}(\eta ,
\rho ,\chi ,S)+\lambda
_{_{15}}\eta ^{\dagger }S^{\dagger }S\eta +
4\lambda _{_{16}}\rho
^{\dagger
}S^{\dagger }S\rho +4\lambda _{_{17}}\chi ^{\dagger }
S^{\dagger }S\chi
\nonumber \\
&&+2\sqrt{2}\lambda _{_{18}}\rho ^{\dagger }S^{\dagger }
\rho \eta +
2\sqrt{2}\lambda_{_{19}}\chi ^{\dagger }S^{\dagger }\chi \eta +
\lambda_{_{20}}S^{\dagger
}S^{\dagger }\eta \eta +h.c.
\label{genpoten}
\end{eqnarray}
where
\[
\begin{array}{c}
\rho ^{\dagger }S^{\dagger }\rho \eta =
\varepsilon _{ijk}\ \rho _l^{\dagger }
S^{li\dagger }\rho ^j\eta ^k~,
\\
\chi ^{\dagger }S^{\dagger }\chi \eta =
\varepsilon _{ijk}\ \chi _l^{\dagger }
S^{li\dagger }\chi ^j\eta ^k~,
\\
S^{\dagger}S^{\dagger }\eta \eta=
\varepsilon _{ijk}\ \varepsilon _{lmn}\
\eta ^k\eta ^nS^{il\dagger
}S^{jm\dagger }.
\end{array}
\]

The triplet Higgs fields $\eta ^o,\rho ^o,$and $\chi ^o$
develop VEVs $v,u,$
and $w$, respectively, as follows
\begin{equation}
\langle \eta \rangle =\left(
\begin{array}{l}
v/\sqrt{2} \\ 0 \\ 0
\end{array}
\right) ,\qquad \langle \rho \rangle =\left(
\begin{array}{l}
0 \\ u/\sqrt{2} \\ 0
\end{array}
\right) ,\qquad \langle \chi \rangle =\left(
\begin{array}{l}
0 \\ 0 \\ w/\sqrt{2}
\end{array}
\right) , \label{VEVs}
\end{equation}
while for the sextet, only the $\sigma _{_2}^o$
field develops a VEV
(a nonzero VEV of $\sigma _{_1}^o$ should provide
a nonzero neutrino
mass which, however, we do not consider here)
\begin{equation}
\langle S\rangle =\left(
\begin{array}{ccc}
0 & 0 & 0 \\ 0 & 0 & v_{_\sigma }/2 \\ 0 & v_{_\sigma
}/2 & 0
\end{array}
\right) . \label{sextetVEV}
\end{equation}
Here, all VEVs are taken to be real. (We are restrained
ourselves the possibility
of CP--violation arising from complex VEVs which has already
been investigated in
detail by D. G. Dumm \cite{tj}). The expansion of
the scalar fields reads
\begin{eqnarray}
\eta ^o=\frac{v}{\sqrt{2}}+\xi _\eta +i\zeta _\eta , & &\quad \rho
^o=\frac{u}{\sqrt{2}}+\xi _\rho +i\zeta_\rho, \;\;\;\;\; \quad
\chi ^o=\frac{w}{\sqrt{2}}+\xi _\chi +i\zeta _\chi ,\nonumber \\ &
&\quad \frac{\sigma _{2}^o}{\sqrt{2}}=\frac{v_{\sigma}}2 + \xi
_\sigma +i\zeta _\sigma , \label{scalars3}
\end{eqnarray}
and
\begin{equation}
\sigma _{1}^o=\xi _\sigma ^{\prime }+i\zeta
_\sigma ^{\prime }.
\label{scalars6}
\end{equation}
Below we call a real part $\xi $ scalar and an
imaginary one $\zeta $
pseudoscalar. In this case the symmetry breaking ladder is
\[
\begin{array}{c}
SU_{C}(3)\otimes SU_{L}(3)\otimes U_{N}(1) \\
\downarrow \ \langle \chi \rangle \\
SU_{C}(3)\otimes SU_{L}(2)\otimes U_{Y}(1) \\
\qquad \quad \quad \downarrow \ \langle \rho \rangle ,
\langle \eta
\rangle
,\langle S\rangle \\
SU_{C}(3)\otimes U_{Q}(1)
\end{array}
\]
where the VEVs satisfy the relation
\begin{equation}
v^{2}+u^{2}+v_{_{\sigma }}^{2}\equiv v_{_{W}}^{2}
\approx (246~~\normalsize{ GeV})^{2},
\label{vev}
\end{equation}
with $v_{_{W}}$ being the standard model VEV. At the first
step of the symmetry
breaking, $\langle \chi \rangle $ generates masses
for exotic quarks and new
heavy gauge bosons. At the subsequent breaking, nonzero
values of $\langle \rho \rangle ,$ $\langle \eta
\rangle $ and $\langle S\rangle $ generates masses
of the familiar quarks and leptons. To keep the
3 - 3 - 1 models consistent with the low-energy phenomenology,
the VEV $\langle \chi
\rangle $ must be large enough in comparison with other VEV's
\[
w\gg v,u,v_{_{\sigma }}.
\]

Due to the requirement the potential to reach a
minimum at the chosen VEV's we obtain the following
constraint equations in the tree--level approximation

\begin{eqnarray}
\mu _{_1}^2 &=&-\lambda _{_1}v^2-\frac{\lambda _{_4}}2
u^2-\frac{\lambda_{_5}}2 w^2-\left(\frac{\lambda_{_{12}}}2-\lambda
_{_{20}}\right)v_{_\sigma }^2+\frac{\lambda
_{_{18}}u^2v_{_\sigma}}{v}-\frac{\lambda_{_{19}}w^2v_{_\sigma
}}{v}-\frac{f_{_1}uw}{v},  \nonumber \\ 
\mu_{_2}^2 &=&-\lambda _{_2}u^2-\frac{\lambda _{_4}}2
v^2-\frac{\lambda_{_6}}2
w^2-\left(\frac{\lambda_{_{13}}}2 +\lambda _{_{16}}
\right)v_{_\sigma}^2 
+2\lambda_{_{18}}vv_{_\sigma}
-\frac{f_{_1}vw}{u}-\frac{f_{_2}v_{_\sigma}w}{u},
\nonumber \\ 
\mu _{_3}^2 &=&-\lambda _{_3}w^2-
\frac{\lambda_{_5}}2 v^2-\frac{\lambda_{_6}}2
u^2-\left(\frac{\lambda_{_{14}}}2+
\lambda _{_{17}}\right)v_{_\sigma
}^2-2\lambda_{_{19}}vv_{_\sigma
}-\frac{f_{_1}vu}{w}
-\frac{f_{_2}v_{_\sigma}u}{w},
\label{constraint}
\\
\mu _{_4}^2 &=&-\left(\lambda _{_{10}}+\frac{\lambda
_{_{11}}}2\right)v_{_\sigma}^2- 
\left(\frac{\lambda_{_{12}}}2-\lambda
_{_{20}}\right)v^2- 
\left(\frac{\lambda _{_{13}}}2+
\lambda_{_{16}}\right)u^2 - 
\left(\frac{\lambda_{_{14}}}2+\lambda
_{_{17}}\right)w^2  \nonumber \\ &&+2\frac{\lambda
_{_{18}}u^2v}{v_{_\sigma
}}-2\frac{\lambda_{_{19}}w^2v}{v_{_\sigma
}}-\frac{f_{_2}wu}{v_{_\sigma }}  \nonumber
\end{eqnarray}
which, in fact, exlude the linear terms in fields from the
potential. The mass matrices, thus, can be calculated, using
\[
M_{ij}^2=\frac{\partial ^2V_{_E}}{\partial
\phi _i\partial \phi _j}
\]
evaluated at the chosen minimum, where
$\phi _i$'s are fields ($\xi,\zeta
,\eta ,\rho ,\chi ,s$).

\section{Higgs mass spectra and physical particles}

Since the $\sigma _{_1}^o$ field does not develop
a VEV, the associated
scalar $\xi _{_\sigma }^{\prime }$ and pseudoscalar
$\zeta _{_\sigma}^{\prime }$ do not mix with other
fields and we have the physical field
$H_{_\sigma }^{\prime }\simeq \xi _{_\sigma }^{\prime }$
with mass
\begin{equation}
m_{H_{_\sigma }^{\prime }}^2=\frac{f_{_2}uw}{v_{_\sigma
}}+\frac{\lambda_{_{11}}}2 v_{_\sigma }^2-(\frac{\lambda
_{_{15}}}2+\lambda _{_{20}})v^2+\lambda_{_{16}}
u^2+\lambda _{_{17}}
w^2-\frac{\lambda_{_{18}}u^2v}{v_{_\sigma
}}+\frac{\lambda _{_{19}}w^2v}{v_{_\sigma }}.
\label{sigmamass}
\end{equation}

In the basis of $\xi _\eta ,\xi _\rho ,\xi _\sigma $ and
$\xi _\chi $ the square
mass matrix, after imposing the constraints
(\ref{constraint}), reads
\begin{equation}
M_{_{4\xi }}^2=\left(
\begin{array}{cccc}
-m_{_{\xi _{_{_\eta }}}}^2 & -m_{_{\xi _{_{_\eta
}}\xi _{_\rho }}}^2 &
-m_{_{\xi _{_{_\eta }}\xi _{_\sigma }}}^2
& -m_{_{\xi _{_{_\eta }}\xi
_{_\chi }}}^2 \\
& -m_{_{\xi _{_\rho }}}^2 & -m_{_{\xi _{_{_\rho
}}\xi _{_\sigma }}}^2 &
-m_{_{\xi _{_{_\rho }}\xi _{_\chi }}}^2 \\
&  & -m_{_{\xi _{_{_\sigma }}}}^2
& -m_{_{\xi _{_{_\sigma }}\xi _{_\chi
}}}^2
\\
&  &  & -m_{_{\xi _{_{_\chi }}}}^2
\end{array}
\right)
\label{scalarmatrix}
\end{equation}
where
\[
m_{_{\xi _{_{_\eta }}}}^2=\frac{f_{_1}uw}{v}
-2\lambda_{_1}v^2-
\frac{\lambda _{_{18}}u^2v_{_\sigma}}{v}+\frac{\lambda
_{_{19}}w^2v_{_\sigma}}{v}~,~~
m_{_{\xi _{_{_\eta }}\xi
_{_\rho
}}}^2=-f_{_1}w-\lambda_{_4}uv+2\lambda
_{_{18}}uv_{_\sigma },
\]
\[
m_{_{\xi _{_\rho }}}^2=\frac
w{u}(f_{_1}v+f_{_2}v_{_\sigma
})-2\lambda_{_2}u^2~,~~
m_{_{\xi _{_{_\eta }}\xi _{_\sigma
}}}^2=-(\lambda
_{_{12}}-2\lambda_{_{20}})vv_{_\sigma} 
+ \lambda_{_{18}}u^2-\lambda _{_{19}}w^2,
\]
\[
m_{_{\xi _{_{_\rho }}\xi _{_\sigma
}}}^2=-f_{_2}w-(\lambda_{_{13}}+
2\lambda_{_{16}})uv_{_\sigma }+
2\lambda _{_{18}}uv~,~~
m_{_{\xi
_{_{_\eta }}\xi _{_\chi
}}}^2=-f_{_1}u-\lambda_{_5}vw-2\lambda
_{_{19}}wv_{_\sigma },
\]
\[
m_{_{\xi _{_{_\rho }}\xi _{_\chi
}}}^2=-f_{_1}v+
f_{_2}v_{_\sigma} -\lambda
_{_6}uw~,~~
m_{_{\xi _{_{_\sigma }}\xi _{_\chi
}}}^2=-f_{_2}u-
(\lambda_{_{14}}+2\lambda
_{_{17}})wv_{_\sigma }
-2\lambda _{_{19}}wv,
\]
and
\[
m_{_{\xi _{_{_\sigma }}}}^2\equiv
\frac{f_{_2}uw}{v_{_\sigma}}-(2\lambda_{_{10}}+\lambda
_{_{11}})v_{_\sigma }^2
-\frac{\lambda_{_{18}}u^2v}{v_{_\sigma }}+\frac{\lambda
_{_{19}}w^2v}{v_{_\sigma }},
\]
\[
m_{_{\xi _{_{_\chi }}}}^2\equiv \frac
u{w}(f_{_1}v+f_{_2}v_{_\sigma})-2\lambda
_{_3}w^2.
\]
As in \cite{ton} we use here the approximation
$\left| f_{_1}\right|,\left| f_{_2}\right| \sim w$
and maintain only terms of the second order in
$w$ in (\ref{scalarmatrix}) (using $w\gg v,u,v_{_\sigma }$).
This immediately gives us one physical field
\begin{equation}
H_{_\chi }\simeq \xi _{_\chi }
\label{khi}
\end{equation}
with a mass
\begin{equation}
m_{_{H_{_{_\chi }}}}^2\simeq -2\lambda _{_3}w^2, \label{khimass}
\end{equation}
and a square mass matrix of $\xi _\eta ,\xi _\rho ,
\xi _\sigma $ mixing
\begin{equation} 
M_{_{3\xi }}^2\simeq w\left( \begin{array}{ccc}-
\frac{f_{_1}u}{v}-
\frac{\lambda _{_{19}}wv_{_\sigma }}{v} 
&f_{_1} & \lambda _{_{19}}w\\
f_{_1} & -
{1\over u}
\left (f_{_1}v+f_{_2}v_{_\sigma }\right)
&f_{_2}\\ \lambda _{_{19}}w&
f_{_2} & -\frac{f_{_2}u}{v_{_\sigma}} 
- \frac{\lambda _{_{19}}wv}{v_{_\sigma }}
\end{array}
\right) . \label{scalarmixing}
\end{equation}
Solving the charateristic equation for the matrix
(\ref{scalarmixing}) we get one massless
field $H_{_1}$ and two physical ones $(H_{_2},H_{_3})$
with masses
\begin{eqnarray}
x_{_{2,3}} &=&-\frac w{2}\left[
\frac{f_{_1}}{uv}(u^2 + v^2)+
\frac{f_2}{uv_{_\sigma
}}(v_{_\sigma }^2+u^2)+
\frac{\lambda_{_{19}}w}{vv_{_\sigma
}}(v^2+v_{_\sigma}^2)\right]  \nonumber \\ & &\pm 
\frac w{2}\left\{ \left[
\frac{f_{_1}}{vu}(v^2+u^2)+
\frac{f_2}{uv_{_\sigma
}}(v_{_\sigma }^2+u^2)+
\frac{\lambda_{_{19}}w}{vv_{_\sigma}}
(v^2+v_{_\sigma}^2)\right]^2\right.  \nonumber \\ & &
\left. -4v_{_w}^2
\left[ {f_{_1}f_{_2}\over vv_{_\sigma}}+
{\lambda_{_{19}}w\over u}
\left({f_{_1}\over v_{_\sigma}}+
\frac{f_{_2}}{v}\right)\right] \right\} ^{1/2}\nonumber \\ 
&\equiv &m_{H_{2,3}}^2
\label{hmass}
\end{eqnarray}
The characteristic equation corresponding to $x_{_{2,3}}$ can 
be given in the following compact form 
\begin{equation}
v\left[F_{_2}(n) + G_{_1}\right] +u\left[F_{_1}(n)F_{_2}(n)
-G_{_1}G_{_2}\right] + 
v_{_\sigma}\left[ F_{_1}(n) + G_{_2}\right]=0, ~~~n,=2,3, 
\label{characeq23}
\end{equation}
where 
\begin{eqnarray}
F_1(i)&=& {u\over v} +
G_1{v_{_\sigma}\over  v} + 
{x_i \over  f_1 w}, \nonumber\\
F_2(i)&=& {u\over v_{_\sigma}}+G_2{v\over v_{_\sigma}}
 + {x_i\over  f_2w}, \nonumber\\
G_1&=& {\lambda_{_{19}}w\over f_1}, \nonumber\\
G_2&=& {\lambda_{_{19}}w\over f_2}.
\label{fg}
\end{eqnarray}

To construct physical fields we now consider the equation
\begin{equation}
\left( M^2-x_i\right) H_i=0,\qquad i=1,2,3,
\label{eigeneq}
\end{equation}
where $M^2$ is a square mass matrix and $H_i\equiv
(H_{i1},H_{i2},H_{i3})^T$. For $M_{_{3\xi }}^2$ we
obtain a system of three equations
\begin{eqnarray}
-\left( \frac{f_{_1}uw}{v}+\frac{\lambda _{_{19}}w^2
v_{_\sigma}}{v}+x_i\right)H_{i1}+
f_{_1}wH_{i2}+
\lambda _{_{19}}w^2H_{i3} &=&0,\nonumber\\
f_{_1}wH_{i1}-\left[ \frac
w{u}(f_{_1}v+f_{_2}v_{_\sigma})+x_i\right]
H_{i2}+
f_{_2}w H_{i3} &=&0, \label{eqsystem}\\
\lambda _{_{19}}w^2H_{i1}
+f_2wH_{i2}
-\left( \frac{f_{_2}uw}{v_{_\sigma }}+\frac{\lambda
_{_{19}}w^2 v}{v_{_\sigma }}+x_i\right)
H_{i3}&=&0.\nonumber
\end{eqnarray}
It is clear that this system of equations is over
defined and can be
reduced to two equations, say, the first and
the last ones. Thus we
have a freedom to suppose
\begin{equation}
H_{i1}=k(i),
\label{hi1}
\end{equation}
where $k(i)$ will be defined by the normalization
of the states. Hence,

\begin{eqnarray}
H_{i2}&=& {F_1(i)F_2(i) - G_1G_2
\over F_2(i) + G_1}k(i)\equiv \Gamma _2(i)~k(i)
\label{hi2}
\end{eqnarray}
and
\begin{eqnarray}
H_{i3}&=& {F_1(i)+  G_2
\over F_2(i)+G_1}k(i)\equiv \Gamma _3(i)~k(i).\label{hi3}
\end{eqnarray}
Then $k(i)$ can be found
\begin{equation}
k(i)=\left[ 1+\Gamma _{_2}^2(i)+\Gamma _{_3}^2(i)
\right] ^{-1/2},
\label{ki}
\end{equation}
by normalizing the states $H_i$ written now in the form
\begin{equation}
H_i=k(i)\left(
\begin{array}{c} 1 \\
\Gamma _{_2}(i) \\
\Gamma _{_3}(i)
\end{array}
\right) \equiv \left(
\begin{array}{c}
H_{i1} \\
H_{i2} \\
H_{i3}
\end{array}
\right) .
\label{hiscalar}
\end{equation}

  In the massless ($x_{_1}=0$) approximation $i=1$ we immediately find
\begin{equation}
H_1=\frac 1{v_{_W}}\left(
\begin{array}{c}
v \\ u \\ v_{_\sigma }
\end{array}
\right) . \label{h1massless}
\end{equation}
In the next approximation (when the $\lambda $'s are
taken into account) the
field $H_{_1}$ acquires a mass. Solving the
characteristic equation for the
exact 3 $\times $ 3 mass matrix $M_{_{3\xi }}^2$
and the $H_{_1}$, namely
\begin{equation}
\left( M_{_{3\xi }}^2-x_{_1}\right) H_{_1}=0,
\label{eqforx1}
\end{equation}
we obtain the following formulas for the $H_{_1}$ mass
\begin{eqnarray}
m_{H_{_1}}^2 &=&x_{_1}\approx 2\lambda
_{_1}v^2+\lambda_{_4}u^2+(\lambda_{_{12}}-2\lambda
_{_{20}})v_{_\sigma }^2-2\lambda _{_{18}}\frac{u^2v_{_\sigma
}}{v}\nonumber 
 \\ &\approx & \lambda _{_4}v^2+2\lambda
_{_2}u^2+(\lambda_{_{13}}+
2\lambda_{_{16}})v_{_\sigma}^2
-4\lambda _{_{18}}vv_{_\sigma } \nonumber \\ 
&\approx &
(\lambda _{_{12}}-2\lambda _{_{20}})v^2+(\lambda_{_{13}}
+2\lambda_{_{16}})u^2+
(2\lambda _{_{10}}+\lambda_{_{11}})v_{_\sigma }^2
\nonumber\\&&
-2\lambda _{_{18}}\frac{u^2v}{v_{_\sigma }}.
\label{h1mass}
\end{eqnarray}

 The coupling constants $\lambda$'s must be chosen
such that the Eqs. (\ref{h1mass}) to be
compatible with the Eq. (\ref{vev}) or,
in the geometric language, we need to find on
the sphere (\ref{vev})
all point(s) ($v$, $u$, $v_{_\sigma}$)
where all the surfaces
(\ref{h1mass}) get together. The simplest
solution of this system
of equations (\ref{vev}) and (\ref{h1mass})
could be found if we
accept the following relation among coupling constants
\begin{equation}
\lambda \approx \lambda _{_1}\approx
{1\over 2}\lambda _{_{12}}-\lambda
_{_{20}}\approx \lambda _{_4}/2\approx
\lambda _{_2}\approx 
({1\over 2}\lambda_{_{13}}+\lambda _{_{16}})\approx 
(\lambda _{_{10}}+{1\over 2}\lambda_{_{11}}).
\label{lambda}
\end{equation}
It would be
\begin{equation}
v\approx v_{_\sigma }\approx
\frac u{\sqrt{2}}\approx {v_{_W}\over 2}\approx
123 ~\normalsize{GeV},
\label{vev2}
\end{equation}
following from
\begin{equation}
\frac{u^2v_{_\sigma }}{v}\approx 2vv_{_\sigma }\approx
\frac{u^2v}{v_{_\sigma }} \equiv \delta^2. \label{vev1}
\end{equation}
The assumption (\ref{lambda}) is justified by
examining the latest
VEV's (\ref{vev2}).  It is easily to see here that 
$\delta^2= {1\over 2}v_{_W}^2$. 
Then the mass of $H_{_1}$
would take the value
\begin{equation}
m_{H_{_1}}^2\approx 2\lambda v_{_W}^2 -2\delta^2 \lambda_{_{18}}=
v_{_W}^2
\left(2\lambda-\lambda_{_{18}}\right), \label{mh1}
\end{equation}
while the eigenstates can be expressed, according
to (\ref{khi}),
(\ref{hiscalar}) and (\ref{mh1}), as follows
\begin{eqnarray}
\left(
\begin{array}{c}
H_1 \\ H_2 \\ H_3
\end{array}
\right) &\approx &\left(
\begin{array}{ccc}
\frac{v}{v_{_W}} & \frac{u}{v_{_W}} &
\frac{v_{_\sigma }}{v_{_W}} \\ H_{21} & H_{22} & H_{23} \\
H_{31} & H_{32} & H_{33}
\end{array}
\right) \left(
\begin{array}{c}
\xi _\eta \\ \xi _\rho \\ \xi _\sigma
\end{array}
\right) , \label{hmatrix}
\\[4mm]
H_\chi &\approx &\xi _\chi . \label{khiscalar}
\end{eqnarray}
Since the matrix
\begin{equation}
A_{_{H\xi }}=\left(
\begin{array}{ccc}
\frac{v}{v_{_W}} &
\frac{u}{v_{_W}} &
\frac{v_{_\sigma }}{v_{_W}} \\ 
H_{21} & H_{22} & H_{23} \\ H_{31} & H_{32} & H_{33}
\end{array}
\right) \equiv \left( A_{_{H\xi }}^{-1}\right) ^T,\qquad \det
A_{_{H\xi }}=1 \label{inversematrix}
\end{equation}
in (\ref{hmatrix}) is an orthonormal matrix $SO(3)$
the relation inverse to (\ref{hmatrix}) and
(\ref{khiscalar}) can easily be found
\begin{eqnarray}
\left(
\begin{array}{c}
\xi _\eta \\ \xi _\rho \\ \xi _\sigma
\end{array}
\right) &\approx &\left(
\begin{array}{ccc}
\frac{v}{v_{_W}} & H_{21} &
H_{31} \\ \frac{u}{v_{_W}} &
H_{22} & H_{32} \\ \frac{v_{_\sigma
}}{v_{_W}} & H_{23}
& H_{33}
\end{array}
\right) \left(
\begin{array}{c}
H_1 \\ H_2 \\ H_3
\end{array}
\right) , \label{etamatrix}
\\[4mm]
\xi _\chi &\approx &H_\chi . \label{khieta}
\end{eqnarray}

Similarly, in the pseudoscalar sector we obtain one
physical field
$\zeta_\sigma \equiv \zeta _\sigma ^{\prime }$ with
a mass equal
to the mass of $H_\sigma ^{\prime }$, and the square
mass matrix of
the $\zeta _\eta ,\zeta
_\rho ,\zeta _\sigma ,\zeta _\chi $ mixing
\begin{equation}
M_{_{4\zeta }}^2=\left(
\begin{array}{cccc}
-m_{_{\zeta _{_{_\eta }}}}^2 & -m_{_{\zeta _{_{_\eta }}
\zeta _{_\rho}}}^2 &
-m_{_{\zeta _{_{_\eta }}\zeta _{_\sigma }}}^2 & -m_{_{
\zeta _{_{_\eta}}\zeta _{_\chi }}}^2 \\
& -m_{_{\zeta _{_\rho }}}^2 & -m_{_{\zeta _{_{_\rho }}
\zeta _{_\sigma}}}^2
& -m_{_{\zeta _{_{_\rho }}\zeta _{_\chi }}}^2 \\
&  & -m_{_{\zeta _{_{_\sigma }}}}^2 & -m_{_{
\zeta _{_{_\sigma }}\zeta_{_\chi }}}^2 \\
&  &  & -m_{_{\zeta _{_{_\chi }}}}^2
\end{array}
\right) .
\label{zetamatrix}
\end{equation}
where
$$m_{_{\zeta _{_{_\eta }}}}^2=\frac{f_{_1}uw}{v}
-\frac{\lambda _{_{18}}u^2v_{_\sigma }}{v}+\frac{\lambda
_{_{19}} w^2v_{_\sigma }}{v}-2\lambda _{_{20}}v_{_\sigma
}^2~,
~~~m_{_{\zeta _{_{_\eta }}\zeta _{_\rho }}}^2=
f_{_1}w ~,$$
$$m_{_{\zeta _{_\rho }}}^2=\frac
w{u}(f_{_1}v+f_{_2} v_{_\sigma })~,
~~m_{_{\zeta _{_{_\eta }}\zeta _{_\sigma }}}^2= \lambda
_{_{18}}u^2-\lambda_{_{19}}w^2 +4\sqrt{2}\lambda
_{_{19}}vv_{_\sigma }~,$$
$$m_{_{\zeta _{_{_\rho }}\zeta _{_\sigma }}}^2
=f_{_2}w~,
~~m_{_{\zeta _{_{_\eta }}\zeta _{_\chi }}}^2
=f_{_1}u~,
~~m_{_{\zeta _{_{_\rho }}\zeta _{_\chi }}}^2
=f_{_1}v +f_{_2}v_{_\sigma}~,
~~m_{_{\zeta _{_{_\sigma }}\zeta _{_\chi }}}^2
=f_{_2}u~,$$
and
$$m_{_{\zeta _{_{_\sigma }}}}^2= \frac{f_{_2}uw} 
{v_{_\sigma
}}-\frac{\lambda _{_{18}}u^2v}{v_{_\sigma }}
+\frac{\lambda _{_{19}}w^2v}{v_{_\sigma }} 
-2\lambda
_{_{20}}v^2~,
~~m_{_{\zeta _{_{_\chi }}}}^2= \frac
u{w}(f_{_1}v +f_{_2}v_{_\sigma }).$$
In the approximation $|f_{_1}|,|f_{_2}|\sim w \gg
v,u,v_{_\sigma }$ we obtain one Goldstone boson $G_1
\approx \zeta _\chi$
and the $\zeta _\eta ,\zeta _\rho ,\zeta _\sigma $ mixing
\begin{equation}
M_{_{3\zeta }}^2=w\left(
\begin{array}{ccc}
-\frac{f_{_1}u}{v}-\frac{\lambda _{_{19}}wv_{_\sigma }}{v} 
&-f_{_1} & \lambda _{_{19}}w \\
-f_{_1} & -\frac 1{u}(f_{_1}v+f_{_2}v_{_\sigma}) 
& -f_{_2}\\
\lambda _{_{19}}w & -f_{_2} 
&-\frac{f_{_2}u}{v_{_\sigma}}-\frac{ \lambda
_{_{19}}wv}{v_{_\sigma }}
\end{array}
\right) .
\label{zetamixing}
\end{equation}
It is clear that the characteristic equation in this case
gives the same roots as in the scalar sector,
but a different set of the eigenstates
(simply, make replacements $H_{i2}\rightarrow -H_{i2}$)
\begin{equation}
\left(
\begin{array}{c}
A_1 \\
A_2 \\
A_3
\end{array}
\right) \approx \left(
\begin{array}{ccc}
\frac{v}{v_{_W}} &
-\frac{u}{v_{_W}} &
\frac{v_{_\sigma }}{v_{_W}} \\ H_{21} & -H_{22} & H_{23}
\\ H_{31} & -H_{32} & H_{33}
\end{array}
\right) \left(
\begin{array}{c}
\zeta _\eta \\
\zeta _\rho \\
\zeta _\sigma
\end{array}
\right)
\label{ascalar}
\end{equation}
or equivalently
\begin{equation}
\left(
\begin{array}{c}
\zeta _\eta \\
\zeta _\rho \\
\zeta _\sigma
\end{array}
\right) \approx \left(
\begin{array}{ccc}
\frac {v}{v_{_W}} & H_{21} &
H_{31} \\ -\frac{u}{v_{_W}}
& -H_{22} & -H_{32} \\ \frac{v_{_\sigma
}}{v_{_W}} & H_{23}
& H_{33}
\end{array}
\right) \left(
\begin{array}{c}
A_1 \\
A_2 \\
A_3
\end{array}
\right) .
\label{zetascalar}
\end{equation}

In the singly charged sector the mixing occurs
in the set of $\eta_1^{+},\rho ^{+},s_1^{+}$
and in the set of $\eta _2^{+},\chi^{+},s_2^{+}$
with the following square mass matrices
\begin{equation}
M_{_{+1}}^2=\left(
\begin{array}{ccc}
-m_{_{\eta _{_{_1}}}}^2 & -m_{_{\rho ^{+}
\eta _{_1}^{-}}}^2 &
-m_{_{s_{_{_1}}^{+}\eta _{_1}^{-}}}^2 \\
& -m_{_{\rho ^{+}\rho ^{-}}}^2 & -m_{_{s_{_{_1}}^{+}
\rho ^{-}}}^2 \\
&  & -m_{_{s_{_1}^{+}s_{_1}^{-}}}^2
\end{array}
\right) ,
\label{1plus1matrix}
\end{equation}
where
\[
m_{_{\eta _{_{_1}}}}^2=\frac{f_{_1}uw}{v} -\frac{\lambda
_{_7}u^2}2-(\frac{\lambda _{_{15}}}4 + \lambda _{_{20}})v_{_\sigma
}^2 -\frac{\lambda _{_{18}}u^2v_{_\sigma }}{v}
+\frac{\lambda_{_{19}}w^2v_{_\sigma }}{v}~,
\]
$$m_{_{\rho ^{+}\eta _{_1}^{-}}}^2 =f_{_1}w
-\frac{\lambda _{_7}uv}2-\lambda_{_{18}}uv_{_\sigma
}~,
~~m_{_{\rho ^{+}\rho ^{-}}}^2=\frac w{u}
(f_{_1}v+f_{_2}v_{_\sigma})-
\frac{\lambda_{_7}v^2}2+ \lambda _{_{16}}v_{_\sigma
}^2-2\lambda _{_{18}} vv_{_\sigma }~,$$
$$m_{_{s_{_{_1}}^{+}\eta _{_1}^{-}}}^2 =\frac{\lambda
_{_{15}}vv_{_\sigma }}{4}+\lambda_{_{19}}w^2
-\lambda _{_{20}}vv_{_\sigma }~,
~~m_{_{s_{_{_1}}^{+}\rho ^{-}}}^2
=-f_{_2}w-\lambda _{_{16}}u
v_{_\sigma}+\lambda _{_{18}}uv~,$$
\[
m_{_{s_{_1}^{+}s_{_1}^{-}}}^2= \frac{f_{_2}uw}
{v_{_\sigma}}+
\frac{\lambda _{_{11}}v_{_\sigma }^2}4-(\frac{\lambda
_{_{15}}}4+\lambda _{_{20}})v^2 +\lambda
_{_{16}}u^2-\frac{\lambda_{_{18}}u^2v} {v_{_\sigma
}}+\frac{\lambda _{_{19}}w^2v}{v_{_\sigma }}~,
\]
and
\begin{equation}
M_{_{+2}}^2=\left(
\begin{array}{ccc}
-m_{_{\eta _{_{_2}}}}^2 & -f_{_1}u+\frac{\lambda
_{_8}wv}2 
- \lambda_{_{19}}wv_{_\sigma }&
({1\over 4}\lambda _{_{15}} + \lambda _{_{20}}) vv_{_\sigma} 
+ \lambda _{_{18}}u^2 \\ 
& -m_{_{\chi ^{+}\chi
^{-}}}^2 & 
f_{_2}u+ \lambda
_{_{17}}wv_{_\sigma}
+ \lambda _{_{19}}wv
\\ &  & -m_{_{s_{_2}^{+}s_{_2}^{-}}}^2
\end{array}
\right) .
\label{1plus2matrix}
\end{equation}
where
\[
m_{_{\eta _{_{_2}}}}^2=\frac{f_{_1}uw}{v} 
- \frac{\lambda_{_8}w^2}2-(\frac{\lambda_{_{15}}}4 
+ \lambda _{_{20}})v_{_\sigma
}^2 -\frac{\lambda _{_{18}}u^2v_{_\sigma }}{v} +
\frac{\lambda_{_{19}}w^2v_{_\sigma }}{v}~,
\]
\[
m_{_{\chi ^{+}\chi ^{-}}}^2=\frac u{w}
(f_{_1}v+f_{_2}v_{_\sigma
})-\frac{\lambda_{_8}v^2}2 + \lambda _{_{17}}v_{_\sigma
}^2
+2\lambda _{_{19}}v v_{_\sigma }~,
\]
\[
m_{_{s_{_2}^{+}s_{_2}^{-}}}^2= \frac{f_{_2}uw}
{v_{_\sigma}}
+\frac{\lambda_{_{11}}v_{_\sigma }^2}4 
- ({\lambda_{_{15}}\over 4}
+\lambda _{_{20}})v^2 
+\lambda _{_{17}}w^2 
-\frac{\lambda_{_{18}}u^2v}{v_{_\sigma }} 
+ \frac{\lambda_{_{19}}w^2v}{v_{_\sigma }}.
\]

Applying the approximation for $M_{_{+2}}^2$ we obtain
one Goldstone boson
$G_{_2}^{+}\approx \chi ^{+}$ and two physical
fields associated with $\eta_{_2}^{+}$
and $s_{_2}^{+}$ with masses
\begin{equation}
m_{_{\eta _{_{_2}}}}^2=\frac{f_1uw}{v}-\frac{\lambda _8
w^2}2 +\frac{\lambda_{_{19}}w^2v_{_\sigma }}{v},\qquad
m_{_{s_{_2}^{+}s_{_2}^{-}}}^2=\frac{f_{_2}uw}
{v_{_\sigma
}}+\lambda _{_{17}}w^2
+\frac{\lambda_{_{19}}w^2v}{v_{_\sigma } }. \label{1chargemass}
\end{equation}
For the $\eta _{_1}^{+},\rho ^{+},s_{_1}^{+}$
mixing, we have
\begin{equation}
M_{_{+1}}^2=w\left(
\begin{array}{ccc}
-\frac{f_{_1}u}{v}-\frac{\lambda _{_{19}}wv_{_\sigma }}{v} &
-f_{_1} & -\lambda _{_{19}}w \\
-f_{_1} & -\frac 1{u}(f_{_1}v+f_{_2}v_{_\sigma}) 
& f_{_2}\\
-\lambda _{_{19}}w & f_{_2} &
-\frac{f_{_2}u}{v_{_\sigma}}
-\frac{ \lambda_{_{19}}wv}{v_{_\sigma }}
\end{array}
\right) .
\label{1plus1mixing}
\end{equation}
As before, the characteristic equation of (2.41)
has the same roots, but the
eigenstates are different and are given by
(necessary replacements:
$H_{i2}\rightarrow -H_{i2},H_{i3}\rightarrow -H_{i3}$)
\begin{equation}
\left(
\begin{array}{c}
h_{_1}^{+} \\
h_{_2}^{+} \\
h_{_3}^{+}
\end{array}
\right) \approx \left(
\begin{array}{ccc}
\frac{v}{v_{_W}} &
-\frac{u}{v_{_W}} &
-\frac{v_{_\sigma }}{v_{_W}} \\ H_{21} & -H_{22} & -H_{23}
\\ H_{31} & -H_{32} & -H_{33}
\end{array}
\right) \left(
\begin{array}{c}
\eta _{_1}^{+} \\
\rho ^{+} \\
s_{_1}^{+}
\end{array}
\right)
\label{chargeh}
\end{equation}
or equivalently
\begin{equation}
\left(
\begin{array}{c}
\eta _{_1}^{+} \\
\rho ^{+} \\
s_{_1}^{+}
\end{array}
\right) \approx \left(
\begin{array}{ccc}
\frac{v}{v_{_W}} & H_{21} &
H_{31} \\ -\frac{u}{v_{_W}}
& -H_{22} & -H_{32} \\ -\frac{v_{_\sigma
}}{v_{_W}} &
-H_{23} & -H_{33}
\end{array}
\right) \left(
\begin{array}{c}
h_{_1}^{+} \\
h_{_2}^{+} \\
h_{_3}^{+}
\end{array}
\right) .
\label{chargefields}
\end{equation}

In the doubly charged sector the mixing occurs up all
states $\rho^{++},s_{_2}^{++},\chi ^{++},s_{_1}^{++}$,
and the square mass matrix is given
\begin{equation}
M_{_{4++}}^2=\left(
\begin{array}{cccc}
-m_{_{\rho ^{++}\rho ^{--}}}^2
& -m_{_{s_{_{_2}}^{++}\rho ^{--}}}^2 &
-m_{_{\chi ^{++}\rho ^{--}}}^2
& -m_{_{s_{_{_1}}^{++}\rho ^{--}}}^2 \\
& -m_{_{s_{_{_2}}^{++}s_{_2}^{--}}}^2
& -m_{_{\chi ^{++}s_{_2}^{--}}}^2
&-m_{_{s_{_{_1}}^{++}s_{_2}^{--}}}^2 \\
&  & -m_{_{\chi ^{++}\chi ^{--}}}^2
& -m_{_{s_{_{_1}}^{++}\chi ^{--}}}^2
\\
&  &  & -m_{_{s_{_{_1}}^{++}s_{_{_1}}^{--}}}^2
\end{array}
\right) ,
\label{2plusmatrix}
\end{equation}
where
$$m_{_{\rho ^{++}\rho ^{--}}}^2=\frac w{u}(f_{_1}v +
f_{_2}v_{_\sigma})-\frac{\lambda _{_9}w^2}2
-4\lambda _{_{18}}vv_{_\sigma }~,$$
$$m_{_{s_{_{_2}}^{++}\rho ^{--}}}^2=
-\sqrt{2}(f_{_2}w+\lambda _{_{16}}uv_{_\sigma
} - \lambda _{_{18}}uv)~,
~~m_{_{\chi ^{++}\rho
^{--}}}^2=f_{_1}v
+f_{_2}v_{_\sigma }-\sqrt{2}\lambda _{_{18}}uw~,$$
$$m_{_{\chi ^{++}s_{_2}^{--}}}^2=\sqrt{2}(- \lambda _{_{17}}w
v_{_\sigma }+ \lambda_{_{19}}wv)~,
~~m_{_{s_{_{_1}}^{++}\rho ^{--}}}^2= 
-\sqrt{2}(\lambda_{_{16}}uv_{_\sigma }
 + \lambda_{_{18}}uv)~,$$
$$m_{_{s_{_{_1}}^{++}s_{_2}^{--}}}^2=-\lambda
_{_{20}}v^2~,
~~m_{_{s_{_{_1}}^{++}\chi ^{--}}}^2=
-\sqrt{2}(f_{_2}u + \lambda_{_{17}}w v_{_\sigma
} + \lambda _{_{19}}wv)~,$$
and
\[
m_{_{s_{_{_2}}^{++}s_{_2}^{--}}}^2= \frac{f_{_2}uw} 
{v_{_\sigma}}+ \lambda _{_{16}}u^2-\lambda _{_{17}}w^2
-\frac{\lambda_{_{18}}u^2v}{v_{_\sigma }} + \frac{\lambda
_{_{19}}w^2v}{v_{_\sigma }}- \lambda _{_{20}}v^2~,
\]
\[
m_{_{\chi ^{++}\chi ^{--}}}^2\equiv \frac
u{w}(f_{_1}v+f_{_2}v_{_\sigma})-
\frac{\lambda _{_9}u^2}2
+4\lambda _{_{19}}vv_{_\sigma }~,
\]
\[
m_{_{s_{_{_1}}^{++}s_{_{_1}}^{--}}}^2\equiv
\frac{f_{_2}uw}{v_{_\sigma }}- \lambda
_{_{16}}u^2+\lambda _{_{17}}w^2 -
\frac{\lambda_{_{18}}u^2v}{v_{_\sigma }}+ \frac{\lambda
_{_{19}}w^2v}{v_{_\sigma }}- \lambda _{_{20}}v^2.
\]

By the same way as considered above we obtain one
Goldstone boson
$G_{_3}^{++}\approx \chi ^{++}$ and one physical
field $s_{_1}^{++}$ with mass
\begin{equation}
m_{_{s_{_{_1}}^{++}s_{_{_1}}^{--}}}^2=
\frac{f_{_2}uw}{v_{_\sigma}}+\lambda _{_{17}}
w^2+\frac{\lambda _{_{19}}w^2v}{v_{_\sigma}},
\end{equation}
and a matrix of $\rho ^{++},s_{_2}^{++}$ mixing
\begin{equation}
M_{_{2++}}^2=w\left(
\begin{array}{cc}
-\frac 1{u}(f_{_1}v+f_{_2}v_{_\sigma})+
\frac{\lambda _{_9}w}2 & \sqrt{2}f_{_2}
\\
\sqrt{2}f_{_2}& -
\frac{f_{_2}u}{v_{_\sigma
}}+\lambda _{_{17}}w-\frac{\lambda
_{_{19}}wv}{v_{_\sigma }}
\end{array}
\right) .
\label{smass}
\end{equation}
Solving the characteristic equation we get two
square masses
\begin{eqnarray}
x_{_{4,5}} &=&\frac w2\left[{\lambda _{_9}w\over 2}
+\lambda_{_{17}}w
-\frac{\lambda _{_{19}}wv}{v_{_\sigma}}
-f_{_2}\left( \frac u{v_{_\sigma }}+\frac{v_{_\sigma
}}u\right) -\frac{f_{_1}v}{u} \right]  \nonumber \\ &&\pm
\frac w2\left\{ \left[{\lambda _{_9}w\over 2}
+\lambda_{_{17}}w
-\frac{\lambda _{_{19}}wv}{v_{_\sigma}}
-f_{_2}\left( \frac u{v_{_\sigma }}+\frac{v_{_\sigma
}}u\right) -\frac{f_{_1}v}{u} \right] ^2\right. \nonumber
\\ &&\left. +\frac{2\sqrt{2}}{uv_{_\sigma }}\left[ (f_{_1}v+f_{_2}
v_{_\sigma})-{\lambda _{_9}uw\over 2}\right]
\left( f_{_2}u-\frac{\lambda _{_{17}}wv_{_\sigma
}}{\sqrt{2}}+\lambda _{_{19}}wv\right) - f_{_2}^2
\right\} ^{1/2} \nonumber
\\ &\equiv &m_{_{d_{_1},d_{_2}}}^2. \label{2plusmixing}
\end{eqnarray}
for two physical fields
\begin{equation}
\left(
\begin{array}{l}
d_{_1}^{++} \\
\\
d_{_2}^{++}
\end{array}
\right) =\left(
\begin{array}{cc}
n_{_4}{1\over \sqrt{2}}\left( \frac u{v_{_\sigma
}}+\frac{x_{_4}}{f_{_2}w}
-\frac{\lambda_{_{17}}w}{f_{_2}}
+\frac{\lambda _{_{19}}wv}{f_{_2}
v_{_\sigma }}\right) &n_{_4} \\  
\\ 
n_{_5}{1\over \sqrt{2}}\left( \frac
u{v_{_\sigma
}}+\frac{x_{_5}}{f_{_2}w}-\frac{\lambda
_{_{17}}w}{f_{_2}}
+\frac{\lambda _{_{19}}wv}{f_{_2}
v_{_\sigma }}\right) & n_{_5}
\end{array}
\right) \left(
\begin{array}{l}
\rho ^{{++}} \\
\\
s_{_2}^{++}
\end{array}
\right) ,
\label{2plusmass}
\end{equation}
corresponding to (3.41) in \cite{tkl}
where
\begin{equation}
n_i=\left[ 1+{1\over 2} \left( \frac u{v_{_\sigma
}}+\frac{x_{_i}}{f_{_2}w}-\frac{\lambda
_{_{17}}w}{f_{_2}} + 
\frac{\lambda _{_{19}}wv}{f_{_2}
v_{_\sigma }}\right) ^2\right] ^{-\frac 12},\qquad (i=4,5),
\label{d2plus}
\end{equation}
is found by normalizing states. Here, using a shorter notation 
\begin{equation}
M_{_{2++}}^2=\left(
\begin{array}{cc}a_{11} & a\\
a& a_{22}
\end{array}
\right) .
\label{smass2}
\end{equation}
of the matrix (\ref{smass})
we, however, can rewrite (\ref{2plusmass}) in 
another way
\begin{equation}
\left(
\begin{array}{l}
d_{_1}^{++} \\
d_{_2}^{++}
\end{array}
\right) =\left(
\begin{array}{cc}
aN_{_4}
&X_{_4}N_{_4}
 \\ aN_{_5}
&X_{_5}N_{_5}
\end{array}
\right) \left(
\begin{array}{l}
\rho ^{{++}} \\
s_{_2}^{++}
\end{array}
\right) ,
\label{2plusmass2}
\end{equation}
where
\begin{equation}
N_i=\left(a^2+X_i^2\right) ^{-\frac 12}
\label{d2plus2}
\end{equation}
and 
\begin{equation}
X_i= {1\over 2}\left[-a_{11}+a_{22}
 \pm \sqrt{(a_{11}-a_{22})^2 
+4a^2}~\right], \qquad (i=4,5).
\end{equation}

\section{Conclusion}

  We have just considered the Higgs sector of the minimal
3 - 3 - 1 model under the most general gauge--invariant
potential
conserving lepton numbers. In comparison with the pevious
paper \cite{tkl} the content of the particles and their
multiplet decomposition structure remain the same but most
of the masses get corrections:\newline

-- in the neutral scalar sector, physical fields are:
$H_{_1},H_{_2},
H_{_3},H_
\sigma ^{\prime }$ and $H_\chi $
\begin{eqnarray}
m_{H_{_1}}^2&\approx& 2\lambda 
v_{_W}^2 -2\delta^2 \lambda_{_{18}}=
v_{_W}^2
\left(2\lambda-\lambda_{_{18}}\right)
,\qquad m_{H_2}^2=x_{_2},\qquad
m_{H_{_3}}^2=x_{_3},  \nonumber\\ 
m_{H_{_\sigma }^{\prime }}^2&=&\frac{f_{_2}uw}{v_{_\sigma
}}+\frac{\lambda_{_{11}}}2 v_{_\sigma }^2-\left(\frac{\lambda
_{_{15}}}2+\lambda _{_{20}}\right)v^2+\lambda_{_{16}}
u^2+\lambda _{_{17}}
w^2-\frac{\lambda_{_{18}}u^2v}{v_{_\sigma
}}+\frac{\lambda _{_{19}}w^2v}{v_{_\sigma }}\nonumber\\
&\approx &
\sqrt{2}f_{_2}w+
{1\over 4}\left(\frac{\lambda_{_{11}}}2 -\frac{\lambda
_{_{15}}}2+2\lambda_{_{16}}-2\lambda_{_{18}}
-\lambda _{_{20}}\right)v_{_W}^2+
\left(\lambda _{_{17}}
+\lambda _{_{19}}\right)w^2,\nonumber\\
m_\chi ^2&\approx& -2\lambda_{_3}w^2,
\label{conclud1}
\end{eqnarray}
where the relation (\ref{vev2}) is used,\\

-- in the neutral pseudoscalar sector, physical fields are:
$A_{_{2}},A_{_{3}},A_{\sigma }$ and two Goldstone bosons:
$G_{_{1}}\approx \zeta
_{\chi }$ and $
G_{_{2}}$ (corresponding to the massless $A_{_{1}}$)
\begin{equation}
m_{A_{_{2}}}^{2}=x_{_{2}},\qquad m_{A_{_{3}}}^{2}=
x_{_{3}},\qquad
m_{A_{_{\sigma }}^{\prime }}^{2}=
m_{H_{_{\sigma }}^{\prime }}^{2},
\end{equation}

-- in the singly charged sector, there are
two Goldstone bosons $
G_{_{3}}=h_{_{1}}^{+},G_{_{4}}^{+}\approx \chi ^{+}$
and three physical
fields : $h_{_{2}}^{+},h_{_{3}}^{+},
\eta _{_{2}}^{+},s_{_{2}}^{+}$ with
masses:
\[
m_{h_{2}^{+}}^{2}=m_{H_{_{2}}}^{2},\qquad
m_{h_{3}^{+}}^{2}=m_{H_{_{3}}}^{2},
\]
\begin{equation}
m_{\eta _{_{2}}}^{2}\approx \frac{f_{_{1}}uw}{v}-
\frac{\lambda _{_{8}}w^2}2 + \frac{ \lambda
_{_{19}}w^{2}v_{_{\sigma }}}{2v},\qquad
m_{_{s_{_{2}}^{+}s_{_{2}}^{-}}}^{2}\approx
\frac{f_{_{2}}uw}{v_{_{ \sigma }}} 
+ \lambda_{_{17}}w^2
+ \frac{\lambda _{_{19}}w^{2}v}{v_{_{\sigma}}},
\end{equation}

-- in the doubly charged sector, there are one Goldstone
($ G_{_{5}}^{++}\approx \chi ^{++}$) and three physical
fields $ d_{_{1}}^{++},d_{_{2}}^{++},s_{_{1}}^{++}$
with masses:
\begin{equation}
m_{d_{1}^{++}}^{2}=x_{_{4}},\qquad m_{d_{2}^{++}}^{2}=
x_{_{5}},\qquad m_{_{s_{_{_{1}}}^{++}s_{_{_{1}}}^{--}}}^{2}=
\frac{f_{_{2}}uw}{v_{_{\sigma}} }+\lambda_{_{17}}w^{2} 
+\frac{\lambda _{_{19}}w^{2}v}{v_{_{\sigma}}}
\equiv m_{_{s_{_{2}}^{+}s_{_{2}}^{-}}}^{2}.
\label{conclud4}
\end{equation}
Eqs. (\ref{conclud1} -- \ref{conclud4}) 
show that $f_{_{1}}$ and $f_{_{2}}$ can
take a definite consistent sign and there are three degenerate
states
$H_{_{2}},A_{_{2}}$ and $h_{_{2}}^{+}$ in mass $x_{_{2}}$,
three degenerate states $H_{_{3}},A_{_{3}}$ and
$h_{_{3}}^{+}$ in mass $x_{_{3}}$ and two degenerate states
$H_{_{\sigma}}^{\prime },A_{_{\sigma }}^{\prime }$ in mass
$m_{H_{_{\sigma }}^{\prime}}^{2}$.

  Combining the assumption (\ref{lambda}) and the positiveness
of the mass squares we get the following bounds for coupling
constants
\begin{equation}
\begin{array}{r}
\lambda \approx \lambda _{_1}\approx
{1\over 2}\lambda _{_{12}}-\lambda
_{_{20}}\approx \lambda _{_4}/2\approx
\lambda _{_2}\approx 
({1\over 2}\lambda_{_{13}}+\lambda _{_{16}})\approx 
(\lambda _{_{10}}+{1\over 2}\lambda_{_{11}})
\stackrel{>}{\sim }0, \\
\lambda _{3}\stackrel{<}{\sim }0.
\label{bound}
\end{array}
\end{equation}
Note that new coupling constants $\lambda _{_{15}}$,
$\lambda _{_{17}}$, $\lambda _{_{18}}$ and $\lambda _{_{19}}$
remain unconstrained by (\ref{lambda}) and (\ref{bound}).
It is worth  mentioning that the system of Eqs. (\ref{vev})
and (\ref{h1mass}) may admit more general solutions with
other coupling constants rather than those constrained by
(\ref{lambda}). This question deserves to be furthermore
investigated.

  In conclusion, the present paper is an extension of previous
investigations \cite{ton,tkl} on the Higgs sector of the minimal
3 - 3 - 1 model with three triplets and one sextet. Under the most
general lepton--number conserving potential the mass spectra and
the multiplet decomposition structre of this sector
are investigated
in detail at tree-level. Due to the fact that most of the scalar
masses get corrections the problem with inconsistent signs of
$f_{_{2}}$ arising in the previuos case \cite{ton, tkl} is solved.
The results of this paper are exposed in a systematic order and
a transparent way allowing them to be easily checked and used in
further studies.
\bigskip

{\bf Acknowledgments}
        
  One of the authors (N.A.K.) would like to thank T. Inami and
Department of Physics, Chuo University, Tokyo, Japan, for warm
hospitality and Nishina foundation for financial support.
H.N.L. would like to thank CNRS for financial support and Theory
Group, LAPP, Annecy, France, for kind hospitality. 
We are thankful to the referee for useful remarks on  
the normalization of the sextet and VEV's.

   This work was supported in part by the Vietnam National Research
Programme on Natural Sciences under grant number KT--04.1.2 and
the Research Program on Natural Science of Hanoi National
University under grant number QT 98.04.


\bigskip
\begin{thebibliography}{99}
\bibitem{ppf} F. Pisano and V. Pleitez, Phys. Rev. {\bf D46},
410 (1992);\\
P. H. Frampton, Phys. Rev. Lett. {\bf 69}, 2889 (1992).
\bibitem{rf} R. Foot,
O. F. Hernandez, F. Pisano, and V. Pleitez, Phys. Rev. {\bf D47},
4158 (1993).
\bibitem{svs}M. Singer, J. W. F. Valle, and J. Schechter, Phys. Rev.
{\bf D22}, 738 (1980).
\bibitem{flt} R. Foot, H. N. Long, and Tuan A. Tran,
 Phys. Rev. {\bf D50}, R34 (1994).
\bibitem{hnl}H. N. Long, Phys. Rev.  {\bf D53}, 437 (1996);
Phys. Rev. {\bf D54}, 4691 (1996).
\bibitem{mpp}J. C. Montero, F. Pisano, and V. Pleitez,
Phys. Rev. {\bf D47}, 2918 (1993).
\bibitem{pq}R. D. Peccei and H. R.  Quinn, Phys. Rev.
Lett. {\bf 38}, 1440 (1977);
Phys. Rev. {\bf D16}, 1791  (1977).
\bibitem{pal}P. B. Pal, Phys. Rev. {\bf D52}, 1659 (1995).
\bibitem{dng}D. Ng, Phys. Rev.  {\bf D49},  4805 (1994).
\bibitem{ton}M. D. Tonasse, Phys. Lett. {\bf B381}, 191 (1996).
\bibitem{tkl}N. T. Anh, N. A. Ky, and H. N. Long, [hep-ph/9810273], 
Int. J. Mod. Phys. {\bf A15}, 283 (2000).
\bibitem{tj}M. B. Tully and G. C. Joshi, Preprint UM--P--98/52,
[hep-ph/9810282]; J. T. Liu
and D. Ng, Phys. Rev. {\bf D50}, 548 (1994); D. G. Dumm,
Int. J. Mod. Phys. {\bf A9}, 887 (1994).
\bibitem{plei} J. Montero, C. de S. Pires and V. Pleitez, Phys. Rev. 
{\bf D60}, 115003 (1999).
\end{thebibliography}
\end{document}